\documentclass[onecolumn,showpacs,preprintnumbers,nofootinbib,amsmath,amssymb]{revtex4}
\usepackage[dvips]{graphics}
\usepackage{color}
\usepackage{pstricks}

\newcommand{\dwall}[1]{\raisebox{-.2mm}{{\bf\large #1}}}

\newcommand{\beq}{\begin{equation}}
\newcommand{\eeq}{\end{equation}}
\newcommand{\bma}{\begin{math}}
\newcommand{\ema}{\end{math}}
\newcommand{\beqa}{\begin{eqnarray}}
\newcommand{\eeqa}{\end{eqnarray}}
\def\opone{\le\textbf{}\textbf{}avevmode\hbox{\small1\kern-3.8pt\normalsize1}}

\newcommand {\be}[1]{
{\marginpar{{\scriptsize\ \\ \ #1}}}
\begin{eqnarray} \mbox{$\label{#1}$} }

\newcommand{\vac}[1]{\ensuremath{\lfloor#1\rfloor}}

\newcommand{\ee}{\end{eqnarray}}
\newcommand{\pref}[1]{(\ref{#1})}

\newcommand{\ul}\underline

\newcommand{\id}{{\mathbf 1}}

\allowdisplaybreaks[4]

\newlength{\cwidth}

\newcommand{\tb}[1]{
\settowidth{\cwidth}{#1}
\addtolength{\cwidth}{-2mm}
\makebox[\cwidth][c]{#1}
}
\newcommand{\tbx}[1]{
\settowidth{\cwidth}{#1}
\addtolength{\cwidth}{-1.4mm}
\makebox[\cwidth][c]{#1}
}

\begin{document}

\title{Degeneracy of non-abelian quantum Hall states on the torus: domain walls and conformal field theory}

\author{Eddy Ardonne$^{1,2}$}
\author{Emil J. Bergholtz$^3$}
\author{Janik Kailasvuori$^4$}
\author{Emma Wikberg$^3$}
\affiliation{$^1$Center for the Physics of Information, California Institute of Technology, Pasadena,
CA 91125, USA\\
$^2$Nordita, Roslagstullsbacken 23, SE-106 91 Stockholm, Sweden \\
$^3$Department of Physics, Stockholm University, AlbaNova University Center, SE-106 91 Stockholm,
Sweden \\
$^4$Institut f\"{u}r Theoretische Physik, Freie Universit\"{a}t Berlin,
Arnimallee 14, 14195 Berlin, Germany}

\date{\today}

\begin{abstract}
We analyze the non-abelian Read-Rezayi quantum Hall states on the torus, where it is natural to employ a mapping of the many-body problem onto a one-dimensional lattice model. On the thin torus---the Tao-Thouless (TT) limit---the interacting many-body problem is exactly solvable. The Read-Rezayi states at filling $\nu=\frac k {kM+2}$ are known to be exact ground states of a local repulsive $k+1$-body interaction, and in the TT limit this is manifested in that all states in the ground state manifold have exactly $k$ particles on any $kM+2$ consecutive sites. For $M\neq 0$ the two-body correlations of these states also imply that there is no more than one particle on $M$ adjacent sites. The fractionally charged quasiparticles and quasiholes appear as domain walls between the ground states, and we show that the number of distinct domain wall patterns gives rise to the nontrivial degeneracies, required by the non-abelian statistics of these states.
In the second part of the paper we consider the quasihole degeneracies from a conformal field theory (CFT) perspective, and show that the counting of the domain wall patterns maps one to one on the CFT counting via the fusion rules.
Moreover we extend the CFT analysis to topologies of higher genus.
\end{abstract}


\maketitle

\section{Introduction}

Microscopic wave functions have ever since Laughlin's original work
\cite{laughlin} back in 1983 been instrumental for the understanding
of the fractional quantum Hall effect (FQHE). At Landau level
filling fraction $\nu=1/q$, $q$ odd, Laughlin's construction shows
why an incompressible quantum liquid, with fractionally charged
excitations, may form to minimize the electron-electron repulsion by
optimizing the short-range correlations as two particles approach
each-other.
The Moore-Read (MR) \cite{mr} and Read-Rezayi (RR) states \cite{readrezayi2}
provide a natural extension of this, where the wave functions vanish as
clusters of $k+1$ particles are formed. The latter has filling fraction
$\nu=k/(kM+2)$, describing fermions (bosons) for $M$ odd (even).

Conformal field theory (CFT) plays a central role in the theory of
the FQHE, as it e.g. describes the edge theory and gives a
method for construction of trial wave functions. The understanding of
this connection was boosted by Moore and Read with their seminal
paper from 1991 \cite{mr}, where they suggested a general CFT-FQHE
connection, and in particular showed that (at least)
some FQHE wave functions can be obtained from correlators  in
certain so called rational CFT's. Not only did they reproduce the
Laughlin wave functions, but also put forward the so called
Moore-Read (aka pfaffian) state which supports non-abelian
excitations. It was first suggested in \cite{greiter} that this state
can describe the enigmatic state observed
\cite{willet} at filling $\nu=5/2$. By now, there is ample (numerical)
evidence that this is indeed the case \cite{morf,rh00,moller}. It is exciting
that the first experimental steps towards determining the nature of
the $\nu=5/2$ quantum Hall state have recently been made \cite{miller07}.

Read and Rezayi \cite{rr07} provided support for the suggestions that the $k=3,M=1$
RR state is underlying the QHE observed at filling $\nu=12/5$, but the amount of 
evidence in this case is not as much as for the $\nu=5/2$ state. However, if this state
does indeed describe
the $\nu=12/5$ state, it would open up the fascinating possibility of topological
quantum computation, as braiding of the non-abelian excitations would be protected
from decoherence by topology and the braid group is rich enough, see e.g.
 Ref. \onlinecite{Freedman}. Recently, another state, with
the same non-abelian structure as the MR state, has been proposed
to describe the $\nu=12/5$ QHE \cite{bs07}.

Bosonic RR states have also been predicted to describe the
state of a rapidly rotating Bose-Einstein condensate, in the regime where the
rotation frequency is so high that the vortex lattice (formed at moderate rotation
rates) melts \cite{cwg01}. This system provides a promising
environment for these states as the extremely local potentials used to motivate
the RR states are more realistic in a dilute atomic bose gas than in
the electronic quantum Hall system.

The Read-Rezayi states can be written as (see \cite{cappelli})
\begin{equation}
\Psi_{\rm RR}=
\mathcal{S} \Big{(}\prod_{{i}_1<{j}_1}^{N/k}(z_{i_1}-z_{j_1})^2\cdots
\prod_{{i_k}<{j_k}}^{N/k}(z_{i_k}-z_{j_k})^2\Big{)}
\prod_{i<j}^{N} (z_i-z_j)^M e^{-\frac 1 4 \sum_i |z_i|^2 } \ ,
\label{wf}
\end{equation}
in the disk geometry. Here $\mathcal{S}$ symmetrizes over possible divisions of the
particles into $k$ sets of equal size. As already discussed, these states are constructed as CFT correlators.
By using the operator product expansions of the fields creating the electrons,
one can show that these wave functions are exact ground states of certain local $k+1$-body interactions.

On a torus, the topological properties of various QH phases will be manifest.
A nice exposition of this can be found in \cite{topdegdum}.
In particular, the one-dimensional nature of a Landau level is explicit on the
torus---a natural set of single-particle states exactly maps the two-dimensional
problem onto a one-dimensional lattice model, where all states will be at least
$q-$fold degenerate for states with filling fraction $\nu = p/q$, and fractionally
charged excitations naturally appear as domain walls between degenerate ground states
\cite{anderson,su1984}. Lately there has been considerable progress in understanding
 various phases of the QH system in terms of the exactly solvable Tao-Thouless (TT) limit (corresponding
 to the thin torus), using this mapping. It has been shown that this limit nicely
 accommodates the gapped hierarchical (abelian) \cite{bk2,bklong,we07,hierarchy},
 multicomponent liquids \cite{seidelyang}, as well as gapless fractions \cite{bk,bk2,bklong}.

In the present manuscript we extend earlier results
\cite{we06,seidel06} obtained for non-abelian FQHE liquids in the TT
limit, by considering the thin torus limit of the Read-Rezayi
states, in the presence of excitations, by counting the number of TT
states on the torus. State counting for non-abelian states has been
considered before, and for the MR state, the complete results (on
the sphere) appeared in \cite{readrezayi}. The first results for the
RR states (again for the sphere), appeared in the original paper of
Read and Rezayi, in which the states were defined \cite{readrezayi2}.
More general results, in terms of recursion relations, were obtained
by Gurarie and Rezayi, \cite{gr00}, by exploiting the results of
\cite{bs99}. Explicit counting formulas were obtained in \cite{a02}.
More recently, Read \cite{read06} constructed an explicit set of
wave functions (building on the results of \cite{aks05}),
re-deriving the explicit counting of \cite{a02} in the process. The
paper \cite{read06} also outlines the state-counting on the torus,
in terms of character formulas. Subsequently, the quasihole states
have been interpreted as Jack polynomials for negative parameter
\cite{bh07a,bh07b}, and the orbital occupation numbers used in these
two papers are the ones we employ in this work. Our analysis however
focusses on the domain walls, which represent the excitations, and
the main result is that the domain walls exactly reproduce the
structure of the fusion rules of the conformal field theory
underlying the Read-Rezayi states.

In particular we show how the number of inequivalent
domain walls accounts for the quasiparticle and quasihole
degeneracies needed for non-abelian statistics of these states. We
also explicitly show that this analysis maps one-to-one on the corresponding CFT
analysis of the bulk states, and generalize the counting rules to
arbitrary genus.

In section \ref{ttlimit}, we will introduce the thin torus limit of
the Read-Rezayi states. States without excitations will be
considered first, giving rise to a set of degenerate ground states.
The elementary excitations are represented as domain walls between
these ground states. The counting of these domain walls will be
presented in section \ref{sec:exdeg}. In section \ref{cft}, we will
revisit this counting from a CFT point of view. We will show that
the counting of the domain walls can be mapped one-to-one on the CFT
counting, which reveals that the domain walls provide a particularly
simple way of representing the fusion rules of CFT.

Some details of the counting are collected in several appendices. Appendix \ref{App:higherM}
deals with the generalization of the results to arbitrary $M$ (we assume $M=0$ in the main text).
In Appendix \ref{smatrix} we explain the connection with the $S$-matrix. Some of the details of
the calculation in section \ref{rrcounting} are collected in Appendix \ref{app:rrcounting}.
Finally, Appendix \ref{app:arbg} deals with the generalization to arbitrary genus.

\section{Tao-Thouless limit}
\label{ttlimit}

In this section we describe the Tao-Thouless limit and calculate degeneracies in this solvable limit.

\subsection{One-dimensional description}

The structure of a single Landau level on a torus was first worked
out by Haldane \cite{Haldane85PRL}. Here we provide a simple version
that makes the one-dimensional nature of a Landau level explicit.

We consider a (flat space) torus with lengths $L_1, L_2$ in the $x-$
and $y-$directions respectively. Consistent boundary conditions can be
enforced when $L_1L_2=2\pi N_\phi$  (in units where $\ell=\sqrt{\frac{\hbar c} {eB}}=1$),
where $N_\phi$ is the number of magnetic flux quanta penetrating the surface. In Landau gauge, $\mathbf{A}=By\mathbf{\hat{x}}$,
the states
\begin{equation}
\label{psik}
\psi_j=\pi^{-1/4}L_1^{-1/2}\sum_m
e^{i(\frac{2\pi}{L_1}j+mL_2)x}
e^{-(y+\frac{2\pi}{L_1}j+mL_2)^2/2},\ \quad j=0,1,...,N_\phi-1 \ ,
\end{equation}
form a basis of one-particle states in the lowest Landau level. $\psi_j$ is quasiperiodic and
centered along the line $y=-2\pi j/L_1$. This maps the Landau level
onto a one-dimensional lattice model with lattice constant $2\pi/ L_1$.
A basis of many-particle states is given by $|n_0,n_1,\dots, n_{N_\phi-1}\rangle$,
where $n_j$ is the number of particles occupying site $j$. Due to periodic boundary conditions, $n_{i+N_\phi} \equiv n_i$. The $N-$particle problem at filling fraction
$\nu=N/N_\phi=p/q$ ($p,q$ coprime) can be shown to be at least $q-$fold degenerate for any translation invariant interaction.

On the thin torus, the overlap between different single particle
states goes to zero, and the interacting many-body problem becomes
solvable \cite{bk2}. For generic interactions, the ground states are
regular lattices and the lowest charged excitations appear as domain
walls between degenerate ground states. This limit has been termed
the Tao-Thouless limit \cite{bk2,bklong} since the exact ground
state (for a repulsive two-body interaction) at $\nu=1/3$ in this
limit coincides with the early attempt to explain the quantum Hall
effect by Tao and Thouless \cite{Tao83}. It is interesting to note
that Anderson, already in 1983, noted that the TT state has a finite
overlap with the Laughlin state and he proposed that one can think
of the TT state as a 'parent' evolving into the Laughlin state as
the interaction is turned on \cite{anderson}.

In the following we will show that Read-Rezayi states have a simple
manifestation in this limit and that the counting of quasiparticle
(and/or quasihole) states reduces to a combinatorially simple problem.

\subsection{Ground states}

Before we study the quasiparticle/quasihole states, we will specify the different
ground state sectors. Ground state will in this paper mean a state without
quasiparticles and quasiholes (which we will collectively call excitations).
(These conventions done here are adapted to the somewhat artificial $k+1$-body
interaction. In the physical situation quasiparticles and quasiholes do not only
come as excitations. They are also a necessary part of the ground state when the
system deviates from an exact filling fraction.)

We search for the thin torus ground states for the $k+1$-body interaction at filling $\nu=\frac{k}{kM+2}$.
For a given $k$ and $M$ in the TT limit, ground states are those that fulfill the rules
\begin{itemize}
\item any $kM+2$ consecutive sites contain exactly $k$ particles
\item the distance between two particles is at least $M$.
\end{itemize}
The first rule is a consequence of the $k+1$-body interaction and
the filling, and the second rule can be understood from the two-body
correlations of the Read-Rezayi states. For $M=0$, the above rules
lead to the ground states (in a basis of occupation numbers of the
sites)
\begin{equation}\label{eq:mzerogs}
|k-l,\;l,\;k-l,\;l,\;k-l,\;l,\cdots\rangle, \ \
l=0,1,\ldots, k
\end{equation}
i.e. they have a pattern with the unit cell $\vac{k-l,\;l}$.
For $k=2$, $M=0$ for example, the possible ground states are $|2020\cdots\rangle$, its translated sector
$|0202 \cdots\rangle$, and $|1111\cdots\rangle$. For $k=3$ and $M=1$ one such ground state is
$|0111001110 \cdots\rangle$. The unit cells of these sectors are $\vac{20}$, $\vac{02}$, $\vac{11}$ and
$\vac{01110}$, respectively. For more examples, see Table~\ref{tablegs}. The sectors are
topologically distinct since there is no local process that can transform one sector
into the other without passing through states of higher energy.

\begin{table}[ht]
\renewcommand{\arraystretch}{1.3}
\centering
\begin{tabular}{|c|c|c|c|}
    \hline
 \ $k$ \ &  \  $M$ \ & TT unit cells & degeneracy   \\
    \hline
    \hline
    1 & 2 & $\vac{100}$, \dots  & 3\\
2 & 0 & $\vac{20}$, $\vac{11}$, \dots  & 3 (1)\\
2& 1 & $\vac{1100}$, $\vac{1010}$, \dots & 6 (2)\\
2& 2 & $\vac{101000}$,  $\vac{100100}$, \dots & 9 (3)\\
3& 0 & $\vac{30}$,  $\vac{21}$, \dots & 4\\
3 & 1 & $\vac{11100}$, $\vac{11010}$, \dots & 10 \\
4& 0 & $\vac{40}$, $\vac{31}$, $\vac{22}$, \dots & 5 (1)\\
    \hline\end{tabular}
\caption{Examples of ground state sectors in the TT limit. The dots
denote that  one should complete with all the possible rigid translations
of the presented unit cells to get the full space of degenerate ground states
and the corresponding degeneracy (for $2N_e/k=0 \bmod 2$) to the right.
The degeneracies for $k$ even and $2N_e/k=1\bmod 2$ are indicated in
brackets and are given by unit cells of the kind $\vac{11}$, $\vac{1010}$, $\vac{22}$ etc, with the reduced periodicity $(kM+2)/2$. }
\label{tablegs}
\end{table}

We can now easily reproduce the well known ground state degeneracy on
the torus, i.e. the number of ground state sectors, by counting the
number of different unit cells. Let us assume that $M=0$ and that
the number of sites $N_\phi$ is even (and hence, $N_e = 0\bmod k$).
In that case, all the ground states (\ref{eq:mzerogs}) can be put on
a torus (periodic boundary conditions), so the degeneracy is $k+1$.
For $k$ even, there is one state which can be put on the torus even
when $N_\phi$ is odd, namely $|k/2, \; k/2, \; k/2, \cdots\rangle$.
However, this is the only possibility, so the degeneracy is one. For
arbitrary $M$ (changing only the center of mass degeneracies),
one recovers the results that the degeneracy is
$(k+1)(kM+2)/2$ when $2N_e/k = 0 \bmod 2$, while for $2N_e/k = 1
\bmod 2$ (only possible for $k$ even) the degeneracy is $(kM+2)/2$.

\subsection{Excitations as domain walls}
Excitations are domain walls between different ground state sectors,
as in $|11\dwall{10}202\ldots\rangle$. At such domain walls the above rules
are not satisfied. An isolated \emph{elementary} excitation is
characterized by one string of $kM+2$ consecutive sites carrying $k+1$
or $k-1$ particles, for quasiparticles or quasiholes respectively.
An example with $k=2$, $M=1$:
\setlength{\fboxsep}{.2mm}
\begin{equation}
\label{eq:wallsearch}
|110011\dwall{0010}1010\cdots\rangle 
\end{equation}
where the only string of $kM+2=4$  sites that has deviating particle content  is the string marked  in boldface. (Compare e.g. with the adjacent strings starting at the $6^{th}$  or the $8^{th}$  site.)
In this example, there is a domain wall between the ground state sectors
$|1100\cdots\rangle$ and $|1010\cdots\rangle$ giving a quasihole. We will throughout the paper highlight the strings of deviating particle content in boldface. 

To be able to compactly  characterize states with domain walls we introduce the
following notation. A ground state sector of arbitrary length will be denoted by its
unit cell in between square brackets. The unit cell is extracted from the $kM+2$ first sites of the ground states.  Thus, for $k=3$, $M=0$ we have
\begin{equation}
\label{eq:unitcellbracket}
\begin{array}{cc}
|030303\cdots \rangle & \rightarrow \vac{03}\\
|121212\cdots \rangle & \rightarrow \vac{12}\\
|212121\cdots \rangle & \rightarrow \vac{21}\\
|303030\cdots \rangle & \rightarrow \vac{30} 
\end{array}
\end{equation}
A state with domain walls, i.e. with different sectors, is written as a
sequence of such unit cells. This is done by comparing each sector with the reference ground states. For $k=3$, $M=0$ this is illustrated for the last two sectors of the example state
$|\mathrm{A}\rangle$ below:
\begin{equation}
\label{eq:2112ex}
\begin{array}{cccccc|c|ccc|c|lcllll}
|\mathrm{A}\rangle =
& \tb{$|21212$} & \tb{\dwall{11}}  & \tb{212121} & \tb{\dwall{20}} &
\tb{3} & \tbx{03} & \tb{03} & \tb{\dwall{02}} & \tb{1} & \tbx{21} & \tb{$21\rangle$} & 
\rightarrow & \tb{\vac{21}} & \tb{\vac{12}} & \tb{\vac{03}} & \tb{\vac{12}} \\

&&&&&&&&&&&&&&&&\\

& \tb{$|03030$} & \tb{30} & \tb{303030} & \tb{30} &
\tb{3} & \tb{03} & \tb{03} & \tb{03} & \tb{0} & \tbx{30} & \tb{$303\cdots\rangle$} & 
\rightarrow & & & \tb{\vac{03}} &\\

& \tb{$|12121$} & \tb{21} & \tb{212121} & \tb{21} &
\tb{2} & \tb{12} & \tb{12} & \tb{12} & \tb{1} & \tbx{21} & \tb{$212\cdots\rangle$} & 
\rightarrow & & & & \tb{\vac{12}}
\ .
\end{array}
\end{equation}

This notation in terms of sectors does not specify the state completely, as is seen by states  $|\mathrm{A}\rangle$ and 
$|\mathrm{B}\rangle$ having the
same notation;
\begin{equation}
\label{eq:AB}
\begin{array}{ccccc}
|\mathrm{A}\rangle & = & |21212\dwall{11}212121\dwall{20}30303\dwall{02}12121\rangle  
& \rightarrow & \vac{21} \vac{12}\vac{03}\vac{12} \\
|\mathrm{B}\rangle &= & |2121\dwall{22}12121212\dwall{13}0303\dwall{02}12121\rangle & 
\rightarrow & \vac{21} \vac{12}\vac{03}\vac{12} \\
\end{array}
\end{equation}
though the first domain wall in $|\mathrm{A}\rangle$ is a quasihole and
the first domain wall in $|\mathrm{B}\rangle$  on the other hand is a quasiparticle.
However, one can make the notation unambiguous (up to the length of the intermediate sectors)
by specifying the charge of each domain wall (i.e. if it corresponds to a quasihole or quasiparticle). However, for the non-abelian structure this ambiguity is immaterial, wherefore the compact
notation is useful for us. Nevertheless, as we will see later, the constraint on the number of
particles does depend on whether we have quasiparticles, quasiholes or a mixture.

It is important to notice that not any pair of sectors would give a
domain wall that corresponds to an elementary excitation. Again, we
stress that an elementary excitation is characterized by a single
string containing $k\pm 1$ particles for quasiparticles and
quasiholes respectively. Taking $k=3,M=0$ as an example, starting
from the sector $|2121\rangle$, the only elementary excitations are
given by the domain walls (given in both notations, with the
quasiholes on the left and the quasiparticles on the right)
\begin{align*}
& |212\dwall{11}21212 \cdots\rangle \rightarrow \vac{21}\vac{12}
&& |2121\dwall{22}1212 \cdots\rangle \rightarrow \vac{21}\vac{12}\\
& |2121\dwall{20}3030 \cdots\rangle \rightarrow \vac{21}\vac{30}
& & |212\dwall{13}03030 \cdots\rangle \rightarrow \vac{21}\vac{30}\ .
\end{align*} 
On the other hand the domain wall $|212\dwall{10}3\ldots\rangle$
(i.e. $\vac{21}\vac{03}\dots$) is not of elementary charge. In general we
have for $M=0$ (for $M\neq 0$, see Appendix \ref{App:higherM}) that
only the following domain walls correspond to elementary
excitations:
\begin{align}
\label{elemex}
& \vac{k-l,l}\vac{k-l-1,l+1} & & \vac{k-l,l}\vac{k-l+1,l-1} \qquad \text{for } 0<l<k \\
& \vac{k,0}\vac{k-1,1} & & \vac{0,k}\vac{1,k-1} \ .
\end{align}

The charge of the elementary excitations, for general $M$, is
$e^*=\pm \frac{e}{kM+2}$ (the charge of the particles is set to
$e$). This can be determined by the Su-Schrieffer counting argument
\cite{Schrieffer}. Here we present an alternative way to derive the
charge of the excitations, which can be applied to general filling
fractions $\nu = \frac{p}{q}$ (given the ground states), and which does not require any particular number of
quasiparticles/holes. There are in total $N_\phi$ strings of $kM+2$
consecutive sites,
 one starting at each of the $N_\phi$ sites. In the absence of excitations,
 the density within each string would be  $k/(k M+2)$ particles per
 site and the total charge of the ground state would be
\[
eN_e= N_\phi \times \frac{ek}{kM+2}\, .
\]
In the presence of  $n_\mathrm{qp}$ quasiparticles and  $n_\mathrm{qh}$ quasiholes,
on the other hand, each excitation contributes with one string of deviating density and one has a total charge
\begin{equation}
\label{eq:totalcharge}
e N_e  =   (N_\phi -n_\mathrm{qp}-n_\mathrm{qh}) \frac{ek}{kM+2}+n_\mathrm{qp}\frac{e(k+1)}{kM+2}+n_\mathrm{qh}\frac{e(k-1)}{kM+2}.
\end{equation}
$(N_\phi -n_\mathrm{qp}-n_\mathrm{qh})$ is the number of strings with the original
ground state density. Rewriting this expression we find
\begin{equation}
\label{eq:totalcharge2}
e N_e  = N_\phi \frac{ek}{kM+2} +(n_\mathrm{qp}-n_\mathrm{qh}) \frac{e}{kM+2},
\end{equation}
from which one reads off the charge of the excitations:
\[
e^*= \pm  \frac{e}{kM+2}\, .
\]

Note that \pref{eq:totalcharge2} determines the number of flux quanta, which is integer.
This gives a constraint on the number of particles, quasiparticles and quasiholes.

\subsection{Degeneracy in the presence of excitations}
\label{sec:exdeg}

We will now calculate the degeneracy of the Read-Rezayi states in the
presence of excitations in the TT limit. For the nontrivial part of
the calculation it is enough to study the bosonic $M=0$ case, because
adding an overall Jastrow factor to the wave function can not change
the degeneracy related to the non-abelian statistics.
The reasons for this from the thin-torus point of view are
explained in Appendix~\ref{App:higherM}.

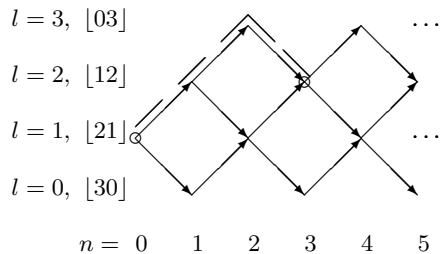
\begin{figure}[ht]
\setlength{\unitlength}{1.5mm}
\begin{picture}(67,32)(-25,-15)

\put(0,0){\circle{1}}
\put(15,5){\circle{1}}

\put(-11,10){$l=3, \ \vac{03}$}
\put(-11,5){$l=2, \  \vac{12}$}
\put(-11,0){$l=1, \  \vac{21}$}
\put(-11,-5){$l=0, \  \vac{30}$}

\multiput(0,1)(3.8,3.8){3}{\line(1,1){2.4}}
\multiput(10,11)(3.1,-3.1){2}{\line(1,-1){2.4}}

\put(0,0){\vector(1,1){5}}
\put(5,5){\vector(1,1){5}}
\put(5,5){\vector(1,-1){5}}
\put(10,0){\vector(1,1){5}}

\put(10,10){\vector(1,-1){5}}
\put(15,5){\vector(1,1){5}}
\put(15,5){\vector(1,-1){5}}

\put(20,0){\vector(1,1){5}}

\put(20,10){\vector(1,-1){5}}
\put(24.5,0){\ldots}
\put(24.5,10){\ldots}

\put(0,0){\vector(1,-1){5}}

\put(5,-5){\vector(1,1){5}}
\put(10,0){\vector(1,-1){5}}

\put(15,-5){\vector(1,1){5}}

\put(20,0){\vector(1,-1){5}}

\put(24.5,0){\ldots}
\put(24.5,10){\ldots}

\put(-5,-10){$n=$}
\put(0,-10){$0$}
\put(5,-10){$1$}
\put(10,-10){$2$}
\put(15,-10){$3$}
\put(20,-10){$4$}
\put(25,-10){$5$}

\end{picture}\\
\caption{An example of a Bratteli diagram, here for those $k=3$, $M=0$,
states that has $\vac{21}$ as starting sector. The arrows indicate which
sectors one can go to from a previous one to get a domain wall of
elementary charge. The circles are added just to guide the eye for
the example discussed in the text. The sequences in  (\ref{eq:AB})
would be represented by the path indicated by the dashed line.}
\label{k4}
\end{figure}

Concentrating on the case $M=0$ from now on, the degeneracy of the
general state containing $n_{qp}$ elementary quasiparticles and
$n_{qh}$ elementary quasiholes is given by the number of distinct
sequences of $n=n_{qp}+n_{qh}$ domain walls as in the examples
to the right in (\ref{eq:AB}). Such sequences can be represented
by Bratteli diagrams as in Fig. \ref{k4}, where the arrows stand for
possible domain walls of elementary charge according to
\pref{elemex}. In Section \ref{cft} we will see that the same
structure appears from the CFT perspective.

Let us consider the simple example of Fig. 1. The $l$ designate the \emph{levels} in the diagram. For $M=0$ each level can be characterized by one sector.  From the diagram we
 infer that for $n=3$ domain walls there is no sequence of
sectors leading back to the starting sector $\vac{21}$. (Note that
ending with the same sector would impose $N_\phi=\textrm{even}$.)
However, there are three paths from $\vac{21}$ at $n=0$ to its translated sector
$\vac{12}$ at $n=3$. (All the allowed paths between the two circles.) By
terminating $\vac{12}$ in the middle of its unit cell (hence
$N_\phi=\textrm{odd}$) as done in  (\ref{eq:2112ex}), periodic
boundary conditions can be fulfilled. Other states with three domain walls and $N_\phi$ odd can be
found by starting with any of the other sectors $\vac{30}$, $\vac{12}$ or
$\vac{03}$ and drawing the same kind of diagram. With the choice $\vac{12}$
the result will of course again be three paths, whereas starting
with $\vac{30}$ and hence terminating with $\vac{03}$ (or vice versa) gives
only one. The torus
degeneracy $td$ for $N_\phi=\textrm{odd}$, $k=3$ and $M=0$ is the sum of all different possibilities for $n=3$ domain walls, resulting in $td=8$. For
$N_\phi=\textrm{even}$, on the other hand, $td=0$ in this case.

These calculations can be formalized by introducing the off-diagonal
$(k+1)\times (k+1)$ adjacency matrix $\mathbf{N}_1$ with
$(N_1)_{ij}=\delta_{i,j+1}+ \delta_{i,j-1}$; $i,j=0,\ldots,k$. It
readily follows that the number of distinct paths $d(k,n,l_1,l_2)$
starting at level $l_1$ and terminating at level $l_2$ via
$n$ domain walls is given by the matrix element of the $n^{\rm
th}$ power of the adjacency matrix;
\begin{equation}\label{eq:ttdegbeforesum}
d(k,n,l_1,l_2)=(N_1^n)_{l_1l_2}.
\end{equation}
The total degeneracy is given by the sum of all allowed paths that
also are compatible with the periodic boundary conditions. When the
number of sites, $N_\phi$, is even, these paths are the ones with $l_1=l_2$,
and when $N_\phi$ is an odd number, they are the ones that  connect level $l_1$ with its complementary level $l_2=k-l_1$. Combining this we find the total degeneracy to be
\begin{equation}
\label{dwc} td(k,M=0,n,\delta) = \sum_{paths} d(k,n,l_1,l_2) =
Tr(\mathbf{N}_1^n \mathbf{B}^\delta).
\end{equation}
Here $\delta=0$ when the sequence returns to its initial level, 
i.e. $l_2=l_1$ ($N_\phi$ even) whereas $\delta=1$ when the sequence terminates at the complementary level
$l_2=k-l_1$  ($N_\phi$ odd). The off-diagonal permutation matrix
$(B)_{ij}=\delta_{i,k-j}$; $i,j=0,\ldots,k$ connects the
complementary levels.

For $M=0$, (\ref{eq:totalcharge2}) yields $N_\phi=(2N_{\rm e} +
n_{\rm qh} - n_{\rm qp})/k$, thus $\delta=(2N_{\rm e} + n_{\rm qh} -
n_{\rm qp})/k$ mod $2$. However, $M$ only affects the center of mass
degeneracy, hence this expression for $\delta$ holds generically (see
also Appendix~\ref{App:higherM}). Thus, for general $M$, we find
\begin{equation}
td(k,M,n,\delta)=\frac{kM+2}{2}
Tr(\mathbf{N}_1^n\mathbf{B}^\delta)\label{td},
\end{equation}
with $\delta = (2N_{\rm e} + n_{\rm qh} - n_{\rm qp})/k  \bmod 2$.  For odd $k$ one can also write 
$\delta = n \bmod 2$ (obvious from inspection of pertinent Bratteli diagrams) and hence replace $\mathbf{B}^\delta$ in (\ref{td}) by $\mathbf{B}^n$ since $\mathbf{B}^2=1$.  For even $k$ one can insert  domain walls only in pairs, which the result (\ref{td}) captures by giving $td=0$ for odd $n$.

We arrived at this formula for the degeneracy of the Read-Rezayi states on the thin
torus by considering the simple picture of domain walls representing the elementary
excitations. As we will see in section \ref{cft}, this formula exactly reproduces the
counting formula one obtains by using more sophisticated conformal field theory methods,
which were used to define the Read-Rezayi states. We will also show how the trace in (\ref{td}) can be evaluated, with the result given in \eqref{tdg1}.

It is however instructive to consider a few simple examples
explicitly, which can be obtained by evaluating the trace.
We will compare these torus degeneracies with the degeneracies on the plane
\footnote{The results for the plane were also given in \cite{sb01},
a nice paper in which the braid properties of the RR quasiholes are calculated.},
namely $pd(k,M,n,\delta=0) = d(k,n,0,0)$ and
$pd(k,M,n,\delta=1) = d(k,n,0,k)$:
\begin{align}
td(2,M,n,\delta)&=\frac{2M+2}{2} \bigl( 2^{n/2+1} +(-1)^{\delta} \delta_{n,0}\bigr) & 
pd(2,M,n,\delta)&= \bigl( 2^{n/2} +(-1)^{\delta} \delta_{n,0}\bigr) \\
td(3,M,n)&= \frac{3M+2}{2} 2\bigl( \mathcal{F}_{n-1}+\mathcal{F}_{n+1} \bigr) & 
pd(3,M,n)&= \mathcal{F}_{n-1} \\ 
td(4,M,n,\delta)&=\frac{4M+2}{2}  \Bigl(2\bigl( 3^{n/2}+
(-1)^{\delta} \bigr) + \delta_{n,0} \Bigr) &
pd(4,M,n,\delta)&= \bigl(
(3^{n/2-1}+(-1)^\delta)/2 + \delta_{n,0}/3 \bigr) \ ,  
\label{tdex}
\end{align}
where it is assumed that $n$ is even for even $k$. $\mathcal{F}_n$
are the Fibonacci numbers,
$\mathcal{F}_n=\mathcal{F}_{n-1}+\mathcal{F}_{n-2}$, with the
initial conditions $\mathcal{F}_0=0$ and $\mathcal{F}_1=1$.

In general, one can write the torus degeneracies in terms of recursion relations.
Taking $k=5$, $M=0$ as an example, one finds the following result:
$td(5,0,n)=td(5,0,n-1)+2 td(5,0,n-2)-td(5,0,n-3)$, with the initial conditions
$td(5,0,0)=6$, $td(5,0,1)=2$ and $td(5,0,2)=10$.

\section{Counting from a conformal field theory perspective}
\label{cft}

In this section, we will show that there is a very close connection between the counting
of the states on the thin torus and conformal field theory. We will do this by performing the counting
in a conformal field theory setting in a rather (perhaps overly) explicit way. This section is written for an audience which is not too familiar with conformal field theory, and would like to understand some of the ideas underlying the state counting by making use of CFT techniques.
As a remark for the experts, one could express the results directly in terms of the modular S-matrix, which diagonalizes the fusion rules \cite{v88}.
However, doing the calculation in a more explicit way nicely reveals the connection with the thin-torus limit. This also has the advantage that we can easily deal with the case $\delta = 1$ (see the previous section), which is more complicated in terms of the $S$-matrix.
In appendix \ref{smatrix} we will make some general remarks about expressing the counting in terms of the $S$-matrix.

\subsection{General remarks}

We will start by explaining the origin of the degeneracy on the
plane\footnote{The results on the sphere are the same, if one
considers localized excitations. In the numerical studies of quantum
Hall states on the sphere, this is not the case if one only fixes
the flux, number of electrons and their interaction. In this case,
the counting is more complicated, see for instance \cite{a02}.} and
torus in general terms, before going into the details of the
specific case at hand.
The conformal field theory, which can be used to describe (or define) quantum Hall states, contains a set of (primary) fields $\phi_a$, which one can think of as the creation operator of particles of type `$a$'.
In order to be a consistent theory, there has to be an `identity' particle (the vacuum). In addition, for each particle $a$, there has to be a dual (or anti-) particle, which we denote by $\bar{a}$.
As a simple example, we consider the description of the $\nu=1/3$ Laughlin state. This theory contains three particle types; $\phi_0$, $\phi_1$ and $\phi_2$, with charges $0$, $e/3$ and $2e/3$.
One can combine two particles with charge
$e/3$ into one particle with charge $2e/3$ by bringing them close together, or in other words, by `fusing' the two particles. In taking an electron completely around any of the three types of particles, one does not pick up any nontrivial phase. In this sense, the electrons are trivial, and correspond to the `identity' sector. Thus, in this theory, charge is defined modulo e. We can now specify the rules stating how particles can be combined, the so called fusion rules. In this case, they are simply given by
$\phi_i\times\phi_j=\phi_{i+j\bmod 3}$.
 
\begin{figure}[ht]
\begin{center}
\psset{unit=1mm,linewidth=.4mm,dimen=middle}
\begin{pspicture}(0,0)(10,10)
\psline(0,0)(10,0)
\psline(5,0)(5,5)
\rput(-2,0){$b$}
\rput(5,7){$a$}
\rput(12,0){$c$}
\end{pspicture}
\end{center}
\caption{Graphical representation of the fusion rule.}
\label{elemfusion}
\end{figure}
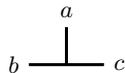

\begin{figure}[ht]
\begin{center}
\psset{unit=1mm,linewidth=.4mm,dimen=middle}
\begin{pspicture}(0,0)(20,10)
\psline(0,0)(20,0)
\psline(5,0)(5,5)
\psline(15,0)(15,5)
\rput(-2,0){$a$}
\rput(5,7){$b$}
\rput(15,7){$c$}
\rput(22,0){$d$}
\rput(10,-2){$e$}
\end{pspicture}
\end{center}
\caption{Fusion of the three particles $a$, $b$ and $c$.}
\label{threefusion}
\end{figure}
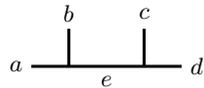

The fusion rules in an abelian theory are of the form $\phi_a \times \phi_b=\phi_{c}$. However, fusing two particles in a  non-abelian theory in general gives more than one possible result. This possibility lies at heart of the degeneracies studied in this paper.
In general, the fusion rules can be written as
\begin{equation}\label{eq:fusionrule}
\phi_a \times \phi_b = \sum_c (N_a)_{b,c} \phi_c \ ,
\end{equation}
where the integer $(N_a)_{b,c}$ is the number of times $\phi_c$ appears in the fusion of the fields $\phi_a$ and $\phi_b$ (note that a more conventional notation would be $N^c_{a,b}$). In this paper, we only consider theories for which $(N_a)_{b,c}=0,1$. In this case, it is easy to represent the fusion rules in a graphical way. The particles are represented by lines, which are labeled by the particle type. The `graph' in Fig. \ref{elemfusion} means that two particles of type $a$ and $b$ can fuse to a particle of type $c$.
The fusion of more than two particles is represented similarly. For instance, fusing three particles $a$,
$b$ and $c$ to a particle of type $d$, namely%
\footnote{One can also first fuse particle $b$ with $c$, and the resulting particle with $a$. Requiring that the two results are the same gives the associativity condition on the fusion matrices.}
\begin{equation}
(\phi_a \times \phi_b) \times \phi_c = \sum_{d,e} (N_b)_{a,e} (N_c)_{e,d} \phi_d \ ,
\end{equation}
is shown in Fig. \ref{threefusion}. Note that in a non-abelian theory, there can be more than one consistent way of labelling this graph, i.e. the label $e$ can take more than one value.

After these general remarks, we will now focus on the Read-Rezayi
states, but  leave the details for the next section. Fusing $n$
quasiholes, which we will denote by $\phi_1$,  leads to
\begin{equation}
\label{eq:multifusion}
\phi_1 \times \phi_1 \times \cdots \times \phi_1 = \sum_{\{a_i\}}
(N_1)_{1,a_1}(N_1)_{a_1,a_2}\cdots (N_1)_{a_{n-2},c} \phi_c=(N_1^{n-1})_{1,c} \phi_c \, .
 \end{equation}
As we will show in the following section, the matrix $\mathbf{N}_1$ describing the fusion of the quasiholes (or quasiparticles) in the Read-Rezayi states is exactly the same matrix describing the possible domain walls between the different sectors. Thus, the rules for domain walls exactly reproduce the fusion rules associated with the Read-Rezayi states! This observation lies at heart of the close connection between the Tao-Thouless states and (the combinatorics of) conformal field theory.

\begin{figure}[ht]
\begin{center}
\psset{unit=1mm,linewidth=.4mm,dimen=middle}
\begin{pspicture}(-5,0)(115,15)
\psline(-5,0)(55,0)
\multips(0,0)(10,0){6}{\psline(0,0)(0,5)}
\multiput(0,0)(10,0){6}{\rput(0,7){$1$}}
\rput(-7,0){$0$}
\rput(57,0){$0$}
\rput(5,-2){$l_1=1$}
\rput(15,-2){$l_2=2$}
\rput(25,-2){$l_3=1$}
\rput(35,-2){$l_4=0$}
\rput(45,-2){$l_5=1$}
\rput(15,0){
\multips(70,0)(10,0){3}{
\psline[linewidth=.2mm]{->}(0,0)(5,5)
\psline[linewidth=.2mm]{->}(5,5)(10,0)
\psline[linewidth=.2mm]{->}(0,10)(5,5)
\psline[linewidth=.2mm]{->}(5,5)(10,10)
\psline[linewidth=.2mm]{->}(0,10)(5,15)
\psline[linewidth=.2mm]{->}(5,15)(10,10)}
\psline[linewidth=.5mm](70,0)(80,10)(90,0)(95,5)(100,0)
\psline[linewidth=.5mm](70,0)(80,10)(90,0)(95,5)(100,0)
\rput[r](69,0){$l=0$}
\rput(68,5){$1$}
\rput(68,10){$2$}
\rput(68,15){$3$}
\rput(65,-5){$n=$}
\rput(70,-5){$0$}
\rput(75,-5){$1$}
\rput(80,-5){$2$}
\rput(85,-5){$3$}
\rput(90,-5){$4$}
\rput(95,-5){$5$}
\rput(100,-5){$6$}
}
\end{pspicture}
\end{center}
\caption{The connection between the labellings of the fusion graph and the
paths on a Bratteli diagram. The labels $\{ l_i \}$ on the fusion graph correspond to the bold path on the Bratteli diagram.}
\label{graphbrat}
\end{figure}
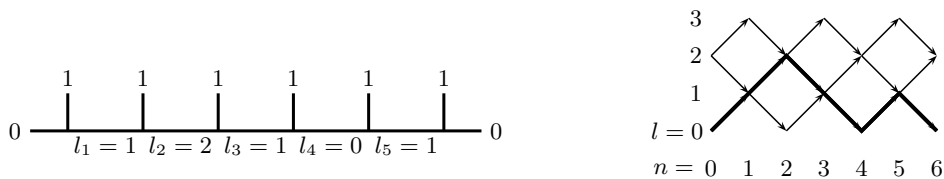

To make this connection more concrete, we consider the degeneracy of
the states with $n$ quasiholes on the plane. In general, one has to
count the number of different ways in which one can fuse all the
fields to the identity. This condition is the non-abelian
generalization of charge neutrality. Taking the case $k=3$, with 
$n=6$ quasiholes as an example,
this degeneracy is given by the number of labellings of the fusion
graph on the left in Fig. \ref{graphbrat}. Each possible labelling
of this graph uniquely corresponds to a path on the Bratteli diagram
on the right.

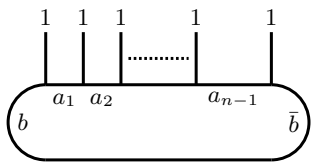
\begin{figure}[ht]
\begin{center}
\psset{unit=1mm,linewidth=.4mm,dimen=middle}
\begin{pspicture}(0,0)(40,20)
\psline(5,0)(35,0)
\psline(5,10)(35,10)
\psarc(5,5){5}{90}{270}
\psarc(35,5){5}{-90}{90}
\psline(5,10)(5,17)
\psline(10,10)(10,17)
\psline(15,10)(15,17)
\psline(25,10)(25,17)
\psline(35,10)(35,17)
\psset{linestyle=dashed}
\psset{dash=1pt 1pt}
\psline(16,13.5)(24,13.5)
\multirput(5,19)(5,0){3}{$1$}
\rput(25,19){$1$}
\rput(35,19){$1$}
\rput(7.5,8){$a_1$}
\rput(12.5,8){$a_2$}
\rput(30,8){$a_{n-1}$}
\rput(2,5){$b$}
\rput(38,5){$\bar{b}$}
\end{pspicture}
\end{center}
\caption{The number of consistent labellings $\{b,\overline{b},a_i\}$ of this graph is the torus degeneracy.}
\label{torgraph}
\end{figure}

On the torus, the situation is slightly different. In this case, one can create a particle-hole pair $b$ and $\bar{b}$ and take one of them, say $b$, around one of the handles of the torus. Then, one could successively fuse all the $n$ quasiholes with the particle $b$, and finally annihilate the resulting particle with $\bar{b}$. Following this logic, the degeneracy on the torus (in the case that $\delta=0$, see the previous section) is given by the number of labellings of the graph of Fig. \ref{torgraph}.
Because all the possible particles $b$ are self-dual (as explained
in the following section), it follows that to obtain the degeneracy
on the torus, one has to count all the paths on the Bratteli diagram,
which begin and end at the same value of $l$, see Fig.
\ref{graphbrat}. Thus, the torus degeneracy (in the case $\delta=0$)
is given by $\sum_{l} (N_1^n)_{l,l} = {\rm Tr}(\mathbf{N}_1^n)$, i.e.
equation \eqref{dwc}. Once again, we see that the structure of the
domain walls in the thin-torus limit precisely reproduces the
results obtained by using the fusion rules. Thus, one can interpret
the domain walls as a very elegant representation of the fusion
rules.

\subsection{Counting of the Read-Rezayi states on the torus}
\label{rrcounting}

We will now move on to the details of the counting of the torus degeneracy of the Read-Rezayi states in the presence of quasiholes (note that we could also consider quasiparticles without additional complication).
We will use the fact that the operators creating the electrons and the quasiholes can be written in terms of fields of the $\mathbf{Z}_k$ parafermion theory (see \cite{zf85}) in combination with a vertex operator, constructed from a compactified chiral boson. In fact, this is the way these states were originally constructed \cite{readrezayi2}.
The vertex operator is purely abelian, and does not contribute to the degeneracy associated with the quasiholes, so we will concentrate on the parafermion part of the theory for now. Of course, the chiral boson will play a role in the center of mass degeneracy later on.

We will think of the $\mathbf{Z}_k$ parafermion theory in terms of the $su(2)_k/u(1)_{2k}$ coset theory. Hence, the parafermion fields $\Phi^l_m$ are labeled by an $su(2)$ label $l$, which takes values $l=0,1,\ldots , k$ and a $u(1)$ label $m$, taking values $m=0,1,\ldots , 2k-1$. Because the $u(1)$ theory is at level $2k$, or radius $\sqrt{2k}$, the label $m$ is defined modulo $2k$, i.e. we have the following identification:
\begin{equation}
\Phi^l_m \equiv \Phi^l_{m+2k} \ .
\end{equation}
Furthermore, from the coset construction, it follows that the labels $l$ and $m$ have to be `compatible', which in the case at hand means that $l+m = 0 \bmod k$. Finally, the requirement that the coset theory is modular yields the following field identification \cite{g89}:
\begin{equation}
\label{fieldident}
\Phi^{l}_{m} \equiv \Phi^{k-l}_{m+k}\ .
\end{equation}
From this it follows that the $\mathbf{Z}_k$ theory has $k(k+1)/2$ primary fields, and
the fusion rules of the theory are determined by the fusion rules of the $su(2)_k$ theory which take the following form:
\begin{equation}
\Phi^{l_1}_{m_1} \times \Phi^{l_2}_{m_3} = \sum_{l_3=|l_1-l_2|}^{\min(l_1+l_2,2k-l_1-l_2)} \Phi^{l_3}_{m_1+m_2}\ , \quad \text{with } l_3=l_1+l_2 \bmod 2 \ .
\end{equation}
The parafermion fields $\psi_i$, with $i=0,1,\ldots,k-1$, are given by $\Phi^{0}_{2i}\equiv \Phi^{k}_{2i-k}$. Their conformal dimension is $h_{\psi_i}= \frac{i(k-i)}{k}$ and they have abelian fusion rules $\psi_{i_1} \times \psi_{i_2} = \psi_{i_1+i_2}$. The spin fields $\sigma_i \equiv \Phi^{i}_{i}$ have scaling dimension $h_{\sigma_i}=\frac{i(k-i)}{2k(k+2)}$.

We can now specify the operators creating the electrons and (elementary) quasiholes for the Read-Rezayi states \cite{readrezayi}. For convenience, we give the explicit form of these operators for the Moore-Read pfaffian state ($k=2$) on the right, by using the operators $\psi$ and $\sigma$ of the Ising CFT.
\begin{align}
V_{\rm el}(z) &= \Phi^{0}_{2}(z) e^{i \sqrt{(kM+2)/k} \varphi_c} (z) & 
V_{\rm el,mr}(z) &= \psi (z) e^{i \sqrt{M+1} \varphi_c} (z) \\
V_{\rm qh}(w) &= \Phi^{1}_{1}(w) e^{i/\sqrt{k(kM+2)} \varphi_c} (w) &
V_{\rm qh,mr}(w) &= \sigma(w) e^{i/\sqrt{4(M+1)} \varphi_c} (w) \ ,
\end{align}
where $\varphi_c$ is a chiral boson, which creates the charge of the electron and quasihole.
From the point of view of the state counting, one can write the operator creating quasiparticles in
the MR state (and similarly in the RR states) as
$V_{\rm qp}(w) = \sigma(w) e^{-i/\sqrt{4(M+1)} \varphi_c} (w)$, though this operator
would not give sensible wave functions. Nevertheless,
this operator can be modified in such a way to give explicit wave functions in the presence of quasiparticles \cite{qpoperator}.

The wave functions of the Read-Rezayi states can be expressed in terms of correlators of the operators $V_{\rm el}$ and $V_{\rm qh}$;
\begin{equation}
\Psi_{\rm RR} = \langle V_{\rm el}(z_1) \cdots V_{\rm el}(z_{N_e})
V_{\rm qh}(w_1) \cdots V_{\rm qh}(w_{n_{\rm qh}}) \mathcal{O}_{\rm bg} \rangle \ ,
\end{equation}
where $\mathcal{O}_{\rm bg}$ is a background operator ensuring $u(1)$-charge neutrality.
The $z_i$ and $w_i$ are the (complex) positions of the electrons and quasiholes respectively.
The form of the wave functions without quasiholes was first given in \cite{readrezayi2}. To obtain the wave functions for an arbitrary number of quasiholes is hard, but for four quasiholes (in which case there are two conformal blocks), they were explicitly calculated in Ref. \onlinecite{as07}.

\begin{figure}[ht]
\begin{center}
\psset{unit=1mm,linewidth=.4mm,dimen=middle}
\begin{pspicture}(0,0)(55,20)
\psline(5,0)(50,0)
\psline(5,10)(50,10)
\psarc(5,5){5}{90}{270}
\psarc(50,5){5}{-90}{90}
\psline(5,10)(5,17)
\psline(10,10)(10,17)
\psline(15,10)(15,17)
\psline(25,10)(25,17)
\psset{linestyle=dashed}
\psline(35,10)(35,17)
\psline(40,10)(40,17)
\psline(50,10)(50,17)
\psset{dash=1pt 1pt}
\psline(16,13.5)(24,13.5)
\psline(41,13.5)(49,13.5)
\rput(5,19){$1$}
\rput(10,19){$2$}
\rput(15,19){$3$}
\rput(25,19){$n_{\rm qh}$}
\rput(35,19){$1$}
\rput(40,19){$2$}
\rput(50,19){$N_e$}
\rput(7.5,8){$a_1$}
\rput(12.5,8){$a_2$}
\rput(29.5,8){$a_{n_{\rm qh}}$}
\rput(2,5){$b$}
\rput(53,5){$\bar{b}$}
\end{pspicture}
\end{center}
\caption{The number of consistent labellings of this graph is the torus degeneracy.}
\label{rrtorus}
\end{figure}
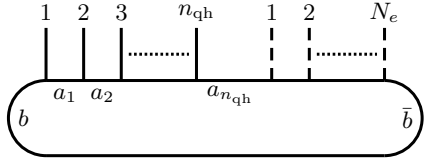

To obtain the degeneracy of the Read-Rezayi states on the torus, in the presence of quasiholes, we have to count the number of consistent labellings of the fusion graph shown in Fig. \ref{rrtorus}. The solid lines numbered $1,\ldots,n_{\rm qh}$ represent the spin field $\Phi^{1}_{1}$, which is associated with the quasiholes. The dashed lines represent the electrons, and correspond to the field $\Phi^{0}_{2}$. The labels $a_i$ and $b$ have to be chosen consistently with the fusion rules. The lines `connecting' the dashed lines representing the electrons do not have a label, because they are completely determined by the fusion rules. Fusing an electron with an arbitrary particle always gives a unique result.
We will concentrate on the bosonic $M=0$ case first. So, we are dealing with the $su(2)_k$ theory,
which has $k+1$ fields that are all self-dual, i.e. $a=\bar{a}$, or in other words, the fusion rules are of the form $a\times\bar{a} = \id + \ldots$. This implies that the label $b$ in Fig. \ref{torgraph} can take $k+1$ different values.

Because we will be considering the insertion of electrons, which are described by descendent fields (namely, of the identity), it turns out that the full counting is most easily done in terms of the parafermion fields. This also allows us to deal with the case $\delta=1$ as well.
In terms of the parafermions, the possible labellings of $b$ are $\Phi^{l}_{l\bmod 2}$, with $l=0,1,\ldots, k$. Modulo the chiral boson factors, which do not affect the counting, these fields are self-dual.
This implies a constraint on the number of quasiholes and electrons. Namely, the $u(1)$ labels of all the inserted fields have to sum to zero, modulo $2k$. Thus, naively one would get the constraint that $2 N_e + n_{\rm qh} = 0 \bmod 2k$. However, by making use of the field identification \eqref{fieldident}, we actually find that consistent labellings are possible when $2 N_e + n_{\rm qh} = 0 \bmod k$, which is precisely the condition that the number of flux quanta is an integer. From this condition, it follows that for $k$ even, $n_{\rm qh}$ also has to be even.

We continue by focussing on the $su(2)$ label $l$ of the parafermion fields. Fusing a field with a quasihole will change this label by one, i.e. it flips the parity:
\begin{equation}
\Phi^1_1 \times \Phi^l_m = \Phi^{l-1}_{m+1} + \Phi^{l+1}_{m+1} \ ,
\end{equation}
where the first and the second field on the right hand side are
present only if $l-1\geq 0$ or $l+1\leq k$, respectively. The
corresponding fusion matrix is given by $(N_1)_{b,c} =
\delta_{b,c-1}+\delta_{b,c+1}$, where $b,c=0,1,\ldots,k$. The only
other way to change the label $l$ is by making use of the field
identification \eqref{fieldident}. This will only change the parity
of the label $l$ in the case that $k$ is odd. Fusing with an
electron, namely $\Phi^0_2$, does not change $l$. Because
all the fields $\Phi^l_{l\bmod 2}$ are self-dual, it follows that
$b=\bar{b}$. Thus, if $b$ takes the value $\Phi^l_{l\bmod 2}$, we
find the condition that after fusing this field with all the $n_{\rm
qh}$ quasiholes, and a possible application of the field
identification \eqref{fieldident}, we should end up with a
parafermion field with the $su(2)$ label $l$. This field should be
fused with the right number of electrons, such that the resulting
field is $\bar{b}$. In this way, we find the consistent labellings
of the graph in Fig. \ref{rrtorus}. Note that the number of electrons is
determined by the relation $2N_e+n_{\rm qh}=0 \bmod 2k$ when the
field identification is not used (this corresponds to the case
$\delta=0$ in the previous section), or $2N_e+n_{\rm qh}=k \bmod
2k$, when it is (i.e., $\delta=1$). Note that there are combinations
of the number of electrons and quasiholes for which neither of these
conditions is satisfied. In those cases, no states exist.

Let us focus on the case $2 N_e + n_{\rm qh} = 0 \bmod 2k$ first, i.e. we do not make use of the field identification. This can occur for $k$ odd and $n_{\rm qh}$ even, or when both $k$ and $(2N_e+n_{\rm qh})/k$ are even. In both cases, we fuse with an even number of quasiholes, from which it follows that there is a parafermion field with label $l$ in the possible fusion outcomes.

We will now explain the connection between the labellings of the graph in Fig. \ref{rrtorus} and the number of paths on the Bratteli diagrams given in section \ref{sec:exdeg}. Let us say we start with a field $\Phi^l_m$. Fusing with a quasihole, or $\Phi^1_1$, can only give two possible results, namely the `neighbouring' fields $\Phi^{l-1}_{m+1}$ and $\Phi^{l+1}_{m+1}$, which are the possibilities for the intermediate label $a_1$. In the Bratteli diagram, this corresponds to the two possible directions, starting from the level $l$ (assuming that $0<l<k$). Repeating this, one finds the correspondance we were after.

Recall that we denote the number of paths on the level $k$ Bratteli diagram, starting at $l_1$ and ending, after fusing $n_{\rm qh}$ quasiholes, at $l_2$ by $d(k,n_{\rm qh},l_1,l_2)$. This number is given by $(N_1^{n_{\rm qh}})_{l_1,l_2}$.
The number of fusion paths in the case at hand is given by $d(k,n_{\rm qh},l,l)$. To obtain the degeneracy on the torus, we have to sum over all possible values of $b$, or, in other words, $l$, so the degeneracy is given by
$\sum_{l=0}^k d(k,n_{\rm qh},l,l)$. In other words, the total number of states is given by the trace of the $n_{\rm qh}$:th power of the fusion matrix $\mathbf{N}_1$, or equivalently, by the sum of all eigenvalues raised to the power $n_{\rm qh}$. In Appendix \ref{app:rrcounting}, it is shown that the eigenvalues in this case are given by $2 \cos \bigl( \frac{(l+1)\pi}{k+2}\bigr)$, with $l=0,1,\ldots,k$. So, in the case that $2N_e+n_{\rm qh}=0\bmod 2k$, we find that the number of $n_{\rm qh}$ quasihole states on the torus is given by
\begin{equation}
\label{even}
\sum_{l=0}^{k} \Bigl( 2 \cos \bigl( \frac{(l+1)\pi}{k+2}\bigr) \Bigr)^{n_{\rm qh}}.
\end{equation}

Let us now consider the case $2N_e + n_{\rm qh} = k \bmod 2k$, where we do make use of the field identification \eqref{fieldident}. This case can occur when both $k$ and $n_{\rm qh}$ are odd or when $k$ is even and $(2N_e +n_{\rm qh})/k$ is odd. When both $k$ and $n_{\rm qh}$ are odd, we need to make use of the field identification, because otherwise no parafermion field with $su(2)$ label $l$ is present in the fusion of the field $b$ and the quasiholes. When $k$ is odd, but $(2N_e+n_{\rm qh})/k$ even, using the field identification does not change the parity of $l$, so after the field identification, the field with label $l$ will be present.
In the Bratteli diagram, the label $l$ occurs at `level' $l$. However, after using the field identification, the `level' at which the label $l$ occurs is $k-l$. Thus, to obtain the torus degeneracy in the case the field identification is used, we need to know the number of paths on the Bratteli diagram which start at $l$, and end at $k-l$, and sum over them, i.e. $\sum_{l=0}^k d(k,n_{\rm qh},l,k-l)$.

In Appendix \ref{app:rrcounting}, we will explain how to calculate this sum and here  we will simply quote the result:
\begin{equation}
\label{odd}
\sum_{l=0}^{k} (-1)^l \Bigl( 2 \cos \bigl( \frac{(l+1)\pi}{k+2}\bigr) \Bigr)^{n_{\rm qh}}.
\end{equation}

The only difference with formula \eqref{even} is the additional sign, which can be explained as follows. In this case, we have an odd number of flux quanta. In transporting one of the particles of the particle-hole pair around one of the handles of the torus, one can pick up a sign, which happens in the case that $b$ corresponds to a field with an odd label $l$.

The results in \eqref{even} and \eqref{odd} are easily combined into one equation, by observing that the additional sign only occurs for the `odd' representations in the case that $N_\phi=(2N_e+n_{\rm qh})/k$ is odd, thus
\begin{equation}
\label{tdM0g1}
{\rm td}(k,M=0,n_{\rm qh},N_e) = \sum_{l=0}^{k} (-1)^{l (2N_e+n_{\rm qh})/k}
\Bigl( 2 \cos \bigl( \frac{(l+1)\pi}{k+2}\bigr) \Bigr)^{n_{\rm qh}} \ .
\end{equation}

To obtain the torus degeneracy for arbitrary $M$, we note that the only thing that changes is the possible values of $b$ in the graph in Fig. \ref{torgraph}. One can show that this gives rise to an additional factor of
$(kM+2)/2$. Thus, we find (taking the possibility of quasiparticles into account as well)
\begin{equation}
\label{tdg1}
{\rm td}(k,M,n_{\rm qh},n_{\rm qp},N_e) =
\Bigl( \frac{kM+2}{2}\Bigr) \sum_{l=0}^{k} (-1)^{l (2N_e+n_{\rm qh}-n_{\rm qp})/k}
\Bigl( 2 \cos \bigl( \frac{(l+1)\pi}{k+2}\bigr) \Bigr)^{n_{\rm qh}+n_{\rm qp}} \ .
\end{equation}

The results presented here for the torus can be generalized to surfaces of arbitrary genus $g$. These results are presented in Appendix \ref{app:arbg}, with the main result being \eqref{eq:arbg}.

\section{Conclusions and Outlook}

We considered the thin torus limit of the Read-Rezayi states. The ground state degeneracy on the
torus is easily obtained in this limit. Elementary excitations correspond to domain walls between the
various ground states. The problem of finding the number of degenerate states in the presence of elementary excitations
translates to the combinatorial problem of finding all the domain walls of elementary charge. This particularly simple picture exactly reproduces the results one obtains by studying the fusion rules of the conformal field theory underlying the Read-Rezayi states. We provided explicit counting formulas for the degeneracy of the RR states on the torus. For completeness, we also give the results for surfaces of arbitrary genus.

The connection between the counting of domain walls and the fusion rules of the $su(2)_k$ conformal field theory describing the Read-Rezayi states can easily be extended to $su(n)_k$. The labels of the ground states correspond one-to-one to the extended labels of the representations of the $su(n)_k$ affine Lie algebra. The domain walls corresponding to quasiholes and -particles are interpreted as fusions with the representations $\omega_1$ and $\omega_{n-1}$ respectively. That the elementary domain walls correctly reproduce these fusion rules is a consequence of the Littlewood-Richardson rule.

{\it Acknowledgements.} We wish to thank A. Karlhede and H. Hansson for fruitful discussions and for comments on this manuscript. J.K. was supported by a grant from the Swedish Research Foundation. 

\appendix

\section{Higher  $M$}
\label{App:higherM}
Increasing  $M$ to   $M\rightarrow M+1$  corresponds to attaching one Jastrow factor, i.e. to push particles away from each-other. In the TT limit one should choose the term in the Jastrow factor that gives the particles the maximal spread.  Up to an  overall translation one has the following one-to-one correspondence: for the transition between $M=0$ and $M=1$ one has ($i,j\geq 0$) 
\begin{equation}
|i_1, \; j_1, \;\; i_2, \; j_2, \cdots \rangle_{M=0} \quad
\leftrightarrow \quad | \underbrace{11\cdots1}_{i_1}0 \;
\underbrace{11\cdots1}_{j_1}0 \; \underbrace{11\cdots1}_{i_2}0
\;\underbrace{11\cdots1}_{j_2}0\cdots \rangle_{M=1} \ .
\end{equation}

The mapping applies for all $M$,  for $M>0$ resulting in the simple rule that every $1$ in the pattern is replaced by $10$ when increasing $M$ by one.  For  example with $k=3$:
\begin{equation}
| 10110 \;\; 10110\cdots \rangle_{M=1} \quad \leftrightarrow \quad
| 10010100 \;\; 10010100\cdots \rangle_{M=2} \ .
\end{equation}

\begin{figure}
\setlength{\unitlength}{1mm}
\begin{picture}(67,32)(15,-15)

\put(-15,13){$l=3$}
\put(-15,6){$l=2$}
\put(-15,-1){$l=1$}
\put(-15,-8){$ l=0$}

\put(-7,-15){$n=$}
\put(0,-1){$\vac{01101}$}
\put(3,-15){0}

\put(12,1){\vector(1,1){5}}
\put(18,6){$\vac{01011}$}
\put(12,-1){\vector(1,-1){5}}
\put(18,-8){$\vac{11100}$}
\put(21,-15){1}

\put(30,8){\vector(1,1){5}}
\put(36,13){$\vac{00111}$}
\put(30,6){\vector(1,-1){5}}
\put(36,-1){$\vac{11010}$}
\put(30,-6){\vector(1,1){5}}
\put(39,-15){2}

\put(48,13){\vector(1,-1){5}}
\put(54,6){$\vac{10110}$}
\put(48,1){\vector(1,1){5}}
\put(54,-8){$\vac{11001}$}
\put(48,-1){\vector(1,-1){5}}
\put(57,-15){3}

\put(66,8){\vector(1,1){5}}
\put(72,13){$\vac{01110}$}
\put(66,6){\vector(1,-1){5}}
\put(72,-1){$\vac{10101}$}
\put(66,-6){\vector(1,1){5}}
\put(75,-15){4}

\put(84,13){\vector(1,-1){5}}
\put(90,6){$\vac{01101}\,\,\dots$}
\put(84,1){\vector(1,1){5}}
\put(90,-8){$\vac{10011}\,\,\dots$}
\put(84,-1){\vector(1,-1){5}}
\put(93,-15){5}

\end{picture}\\
\caption{A Bratteli diagram for $k=3$ and $M=1$ for sequences of quasiholes. Note how the ground states within each level are shuffled around among translated siblings. Not also that as for $M=0$, each level $l_1$ has a complementary level $l_2=k-l_1$ with translated siblings. For sequences of quasiparticles the structure of the diagram will be the same, but the sectors will follow each-other in a different order.}
\label{fig:bratelli01101}
\end{figure}
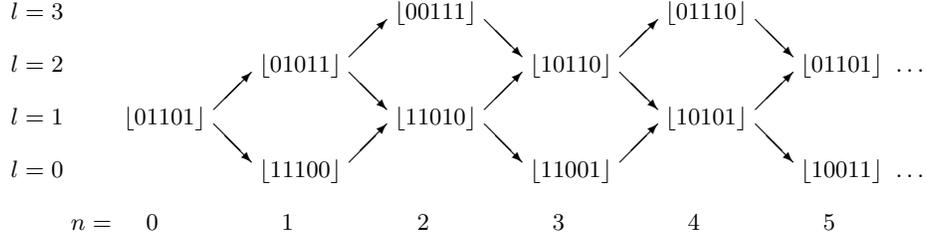

In such a mapping domain walls of elementary charge are mapped to domain walls of elementary charge.  Of course the
absolute value of the charge of the domain wall will
change. Hence, every
state with quasiparticles and quasiholes for some higher $M$ can (up
to an overall translation) be one-to-one mapped to an $M=0$ state
with the same sequence of quasiparticles and quasiholes.  The only change in the degeneracy is that the center of mass part of the degeneracy---related to the overall
translations of a state---goes from $kM+2$ to $2$.
The Bratteli diagrams will have the same topological structure
(compare Fig.~\ref{fig:bratelli01101} and Fig.~\ref{k4}). 
A state keeps the same path in the diagram under the mapping and consequently the value of
$\delta$ does not change. Hence the non-abelian
part of the degeneracy counting will be the same. Note that for
$M>0$ the sectors within a level get shuffled around with
translated siblings and therefore, in contrast to $M=0$, the levels
can no longer be labeled by specific unit cells. Note also that for
$M>0$ a domain wall has no longer the quasiparticle/quasihole
ambiguity. The domain wall $[01101][01110]$ can only be an
elementary quasiparticle, not also an elementary quasihole. This
means that for quasiparticles, the sectors  within each level  in a
Bratteli diagram get shuffled around in a different order compared
to the same Bratteli diagram for quasiholes.

As for $M=0$, the degeneracy formula (\ref{td}) requires $\delta=0$ when the sequences have to  return to the initial level in the diagram  ($l_2=l_1$), whereas $\delta=1$ is needed for the sequences which have to end with the complementary level ($l_2=k-l_1$).
Because increasing $M$ for a given state neither changes the
structure of the state in terms of whether one needs $\delta=0$ or
$\delta=1$, nor changes the number of particles,
quasiparticles and quasiholes, respectively, we have that  $\delta =
(N_{\rm e} + n_{\rm qh} - n_{\rm qp})/k \bmod 2$ must apply also for
$M\neq 0$.  On the other hand, the  equation $\delta= N_\phi \bmod
2$ applying for $M=0$ changes to
$\delta= N_\phi-MN_e \bmod 2$.

\section{Using the $S$-matrix to count conformal blocks}
\label{smatrix}

The material presented in this section is standard (see, for instance, \cite{mooreseiberg}), but is included for completeness.
The Verlinde formula relates the fusion rules of the modular $S$-matrix in the following way: Let
$(N_a)_{b,c}$ denote the number of times the operator $\phi_c$ appears in the fusion product of $\phi_a$ and $\phi_b$, and $S_{a,b}$ the modular $S$-matrix. Then,
\begin{equation}
\label{verlindeform}
(N_a)_{b,c} = \sum_{d} \frac{S_{b,d}^{\vphantom{*}}S_{a,d}^{\vphantom{*}}S_{c,d}^{*}}{S_{\mathbf{1},d}} \ ,
\end{equation}
where the sum is over all fields, and $\mathbf{1}$ denotes the identity field. In other words, the $S$-matrix diagonalizes all the fusion matrices $\mathbf{N}_{a}$ simultaneously; the eigenvalues of $\mathbf{N}_a$ are
$(\lambda_{a})_d=S_{a,d}/S_{\mathbf{1},d}$, where $d$ is any primary field;
\begin{equation}
\sum_{b,c} S^*_{b,d} (N_a)_{b,c} S_{c,e} = \frac{S_{a,d}}{S_{\id,d}} \delta_{d,e} \ .
\end{equation}
Note that $S$ is unitary and symmetric.

\begin{figure}[ht]
\begin{center}
\psset{unit=1mm,linewidth=.4mm,dimen=middle}
\begin{pspicture}(0,0)(60,10)
\psline(0,0)(28,0)
\psset{dash=1pt 1pt}
\psline[linestyle=dashed](30,0)(35,0)
\psline(37,0)(55,0)
\psline(5,0)(5,5)
\psline(15,0)(15,5)
\psline(25,0)(25,5)
\psline(40,0)(40,5)
\psline(50,0)(50,5)
\rput(-2,0){$\mathbf{1}$}
\rput(57,0){$\mathbf{1}$}
\rput(5,7){$i_1$}
\rput(15,7){$i_2$}
\rput(25,7){$i_3$}
\rput(40,7){$i_{n-1}$}
\rput(50,7){$i_n$}
\rput(10,-2){$a_1$}
\rput(20,-2){$a_2$}
\rput(28,-2){$a_3$}
\rput(45,-2){$a_{n-1}$}
\end{pspicture}
\end{center}
\caption{The number of consistent labellings $\{a_i\}$ of this graph gives the number of conformal blocks on the plane.}
\label{confblocks}
\end{figure}
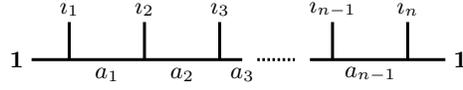

First, we will apply this result to count the number of conformal blocks of the fields $\phi_i$, $i=1,\ldots,n$ on the plane. This number is given by the number of labellings, consistent with the fusion rules, of the graph given in Fig. \ref{confblocks}, namely
$$\#_{g=0}=\sum_{\{ a_i \}} (N_{i_1})_{\mathbf{1},a_1} (N_{i_2})_{a_1,a_2}  \cdots
(N_{i_{n-1}}){a_{n-2},a_{n-1}} (N_{i_n})_{a_{n-1},\mathbf{1}} \ .$$
Inserting the Verlinde formula \eqref{verlindeform}, and first performing the sum over the $a_i$, followed by the sums coming from \eqref{verlindeform}, we obtain
\begin{equation}
\label{sphereblocks}
\#_{g=0}=\sum_a \frac{S_{i_1,a}S_{i_2,a}\cdots S_{i_n,a}}{(S_{\mathbf{1},a})^{n-2}}\ .
\end{equation}

\begin{figure}[ht]
\begin{center}
\psset{unit=1mm,linewidth=.4mm,dimen=middle}
\begin{pspicture}(-2,0)(80,8)
\rput(-2,0){$\mathbf{1}$}
\psline(0,0)(43,0)
\psarc(12.5,0){7.5}{0}{180}
\rput(-2,0){$\mathbf{1}$}
\rput(12.5,-2){$a_1$}
\rput(8,2.5){$b_1$}
\rput(18,2.5){$\overline{b}_1$}
\rput(22.5,-2){$a_2$}
\psarc(32.5,0){7.5}{0}{180}
\rput(32.5,-2){$a_3$}
\rput(28,2.5){$b_2$}
\rput(38,2.5){$\overline{b}_2$}
\psline[linestyle=dashed,dash=1pt 1pt](45,0)(50,0)
\psline(52,0)(75,0)
\rput(77,0){$\mathbf{1}$}
\psarc(62.5,0){7.5}{0}{180}
\rput(62.5,-2){$a_{2g-1}$}
\rput(58,2.5){$b_{g}$}
\rput(68,2.5){$\overline{b}_{g}$}
\end{pspicture}
\end{center}
\caption{The number of consistent labellings of this graph gives the degeneracy of a genus $g$ surface.}
\label{arbggraph}
\end{figure}
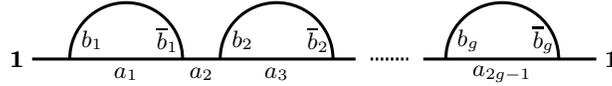
We can now count the number of labellings on the graph in Fig. \ref{arbggraph}, which will give the number of states on the torus, if no fields (or quasiholes in our case) are present. We will make use of the result \eqref{sphereblocks} by choosing $i_{2j-1}=b_j$, $i_{2j}=\overline{b}_j$, for $j=1,2,\ldots,g$, and performing a sum over all possible values of $b_j$.  This leads to the following result:
\begin{equation}
\label{arbg}
\sum_a \bigl( \frac{1}{S_{\id,a}}\bigr)^{2g-2} \ .
\end{equation}

The genus $g$ generalization of \eqref{sphereblocks} is given by
\begin{equation}
\#_{g} = \sum_{a}  S_{i_1,a}\cdots S_{i_n,a} (S_{\id,a})^{2-n-2g} \ ,
\end{equation}
as follows from gluing the graphs in Fig. \ref{confblocks} and Fig. \ref{arbggraph} by summing over all
intermediate states and making use of the unitarity of the $S$-matrix.

\section{Some details of the counting in section \ref{rrcounting}}
\label{app:rrcounting}

In this appendix, we will describe how to obtain the result \eqref{odd}, by making use of the
results in the previous appendix.
We start by first calculating the general result for the number of paths on the Bratteli diagram
$d(k,n_{\rm qh},l_1,l_2)$. In terms of the fusion matrix $\mathbf{N}_1$ (recall that this matrix has the components
$(N_1)_{i,j} = \delta_{i,j+1}+\delta_{i,j-1}$, for $i,j=0,1,\ldots,k$), we have
\begin{equation}
d(k,n_{\rm qh},l_1,l_2) = (N_1^{n_{\rm qh}})_{l_1,l_2} =
\sum_{l=0}^{k} S_{l_1,l} \bigl( \frac{S_{1,l}}{S_{0,l}}\bigr)^{n_{\rm qh}} S_{l_2,l} \ ,
\end{equation}
where the modular $S$-matrix for $su(2)_k$ is given by
(see, for instance, \cite{dms})
\begin{equation}
\label{su2ksmat}
S_{l_1,l_2} = \sqrt{\frac{2}{k+2}} \sin \bigl( \frac{(l_1+1)(l_2+1)\pi}{k+2} \bigr) \ .
\end{equation}
Thus, we obtain
\begin{equation}
\label{degplane}
d(k,n_{\rm qh},l_1,l_2) = \frac{2}{k+2} \sum_{l=0}^{k}
\sin \bigl( \frac{(l+1)(l_1+1)\pi}{k+2} \bigr)
\sin \bigl( \frac{(l+1)(l_2+1)\pi}{k+2} \bigr)
\Bigl( 2 \cos \bigl( \frac{(l+1)\pi}{k+2} \bigr) \Bigr)^{n_{\rm qh}} \ .
\end{equation}
Note that the eigenvalues of $\mathbf{N}_1$ are given by $S_{1,l}/S_{0,l}= 2 \cos \bigl(\frac{(l+1)\pi}{k+2}\bigr)$,
which is a consequence of the Verlinde formula.

We can now perform the sum $\sum_{l=0}^{k} d(k,n_{\rm qh},l,k-l)$ explicitly by making use of
$$
\sin \bigl( \frac{(l+1)(k-l_1+1)\pi}{k+2} \bigr) = (-1)^l \sin \bigl( \frac{(l+1)(l_1+1)\pi}{k+2} \bigr)
$$
and the unitarity of the $S$-matrix. This gives the result stated in the main text \eqref{odd}:
\begin{equation}
\label{app:odd}
\sum_{l=0}^{k} (-1)^l \Bigl( 2 \cos \bigl( \frac{(l+1)\pi}{k+2}\bigr) \Bigr)^{n_{\rm qh}}.
\end{equation}

Note that one can also obtain the eigenvalues of the matrices $\mathbf{N}_1$ by observing that, as a
function of $k$, the characteristic polynomials satisfy a recursion relation, which is the same
as the recursion relation for the Chebyshev polynomials. The zeros of these polynomials
are indeed the eigenvalues we quote above.

\section{Counting results for arbitrary genus}
\label{app:arbg}

For completeness, we give the state counting for arbitrary genus $g$.
By making use of the results in Appendix \ref{smatrix} and \ref{app:rrcounting}, we find
\begin{equation}
\label{eq:arbg}
{\rm td}(k,M,g,n_{\rm qh},n_{\rm qp},N_e) = \Bigl( \frac{kM+2}{2}\Bigr)^{g}
\sum_{l=0}^{k} (-1)^{l (2N_e+n_{\rm qh}-n_{\rm qp})/k}
\Bigl( 2 \cos \bigl( \frac{(l+1)\pi}{k+2}\bigr) \Bigr)^{n_{\rm qh}+n_{\rm qp}}
\Bigl( \frac{k+2}{2\sin\bigl(\frac{(l+1)\pi}{k+2} \bigr)^2}\Bigr)^{g-1} \ .
\end{equation}

Specializing to the case $g=1,n_{\rm qh}=0$, we find that when $N_e = 0 \bmod k$, the degeneracy is given by $(k+1)(kM+2)/2$. When $N_e = k/2 \bmod k$, which only occurs for $k$ even, we find a degeneracy of $(kM+2)/2$. Some other simple results%
\footnote{For $k=2,M=0$, these are known as the boundary condition sectors, or spin structures, of the
genus $g$ torus, as explained in \cite{readgreen}.},
in the absence of quasiholes, are
\begin{align}
{\rm td}(2,M,g,0,N_e) &= ((2M+2)/2)^g 2^{g-1}(2^{g}+(-1)^{N_e}),\\
{\rm td}(3,M,g,0,N_e=0 \bmod 3) &= ((3M+2)/2)^g 2 ((5+\sqrt{5})^{g-1}+(5-\sqrt{5})^{g-1}),\\
{\rm td}(4,M,g,0,N_e=0 \bmod 2) &= ((4M+2)/2)^g (3^{g-1}+(-1)^{N_e/2}2\; 4^{g-1}+2 \; 12^{g-1}).
\end{align}

In the case $g=0$, we reproduce the result that the degeneracy is given by the number of paths on the Bratteli diagram, namely $d(k,n_{\rm qh},0,0)$, or $d(k,n_{\rm qh},0,k)$, \eqref{degplane}, for
$(2N_e+n_{\rm qh})/k = 0 \bmod 2$ or $(2N_e+n_{\rm qh})/k = 1 \bmod 2$ respectively.

\end{document}